\documentclass[10pt]{iopart}
\usepackage{amssymb,graphicx}

\begin{document}
\title[Higher-order glass-transition singularities]{Higher-order glass-transition singularities 
in systems with short-ranged attractive potentials}
\author{W. G\"otze and M. Sperl}
\address{Physik-Department,
Technische Universit\"at M\"unchen, 85747 Garching, Germany}

\begin{abstract}
Within the mode-coupling theory for the evolution of structural relaxation, the $A_4$ glass-transition
singularities are identified for systems of particles interacting with a hard-sphere repulsion 
complemented  by different short-ranged potentials: Baxter's singular potential regularized by a 
large-wave-vector cutoff, a model for the Asakura-Oosawa depletion attraction, a triangular potential,
a Yukawa attraction, and a square-well potential. The regular potentials yield critical packing fractions, 
critical Debye-Waller factors and critical amplitudes very close to each other. The elastic moduli and
the particle's localization lengths for corresponding states of the Yukawa system and the square-well system 
may differ by up to 20\% and 10\%, respectively.
\end{abstract}



\section{Introduction}\label{sec:intro}

In this paper, some results shall be presented for the liquid-glass-transition diagrams of 
simple systems as obtained within the mode-coupling theory for the evolution of structural
relaxation (MCT) \cite{Goetze1991b}. This theory is based on closed non-linear equations of 
motion for the normalized autocorrelation functions for density fluctuations with wave-vector
modulus $q=|\vec{q}|$, $\phi_q(t)$. Liquid states are characterized by solutions vanishing 
for large times $t$, $\phi_q(t\rightarrow\infty)=0$. Glasses are characterized by solutions
whose long-time limits are the non-vanishing Debye-Waller factors of the arrested 
amorphous structure $f_q=\phi_q(t\rightarrow\infty)$, $0<f_q\leq 1$. The $f_q$ obey the set
of implicit equations
\numparts\label{eq:F}
\begin{equation}\label{eq:Feq}
\frac{f_q}{1-f_q} = {\cal F}_q[f_k]\,.
\end{equation}
The mode-coupling functional ${\cal F}_q$ reads
\begin{equation}\label{eq:Fdef}
{\cal F}_q[f_k] = \sum_{\vec{k} + \vec{p} = \vec{q} } V(\vec{q},\vec{k},\vec{p}) f_k f_p
\end{equation}
\endnumparts
with coupling coefficients
\begin{equation}\label{eq:vertex}
V(\vec{q},\vec{k},\vec{p}) = \rho S_q S_k S_p \{[\vec{k}c_k+\vec{p}c_p]\vec{q}/q\}^2/(2q^2)\,.
\end{equation}
Here, $S_q$ and $c_q$ are the structure factor and the direct correlation function,
respectively, related
by the Ornstein-Zernicke equation $S_q = 1/[1- \rho c_q]$. The structure factor depends on 
control parameters like the particle density $\rho$, the temperature $T$, and parameters
specifying the dependence of the interaction potential $U(r)$ on the interparticle distance
$r$. Let us combine the set of control parameters considered to a control-parameter vector
${\mathbf V}$. The discussion will be restricted to parameter regions where $S_q$, and hence 
${\cal F}_q$, depend smoothly on ${\mathbf V}$. The Debye-Waller factor $f_q$ for a given
${\mathbf V}$ is distinguished from possible other solutions of 
Equations~(\ref{eq:F},\ref{eq:vertex}) for the same ${\mathbf V}$, say $\tilde{f}_q$, by the
maximum property: $f_q\geq \tilde{f}_q$ for all $q$.

For almost all control parameters, the  Debye-Waller factors $f_q$ depend smoothly on 
${\mathbf V}$. The exceptional points are referred to as glass-transition singularities
${\mathbf V}^c$. These critical points of Equations~(\ref{eq:F},\ref{eq:vertex})
are bifurcation singularities of the cuspoid family
$A_l,\,l=2,3,\dots$ An $A_l$ bifurcation describes a topologically stable singularity that
is equivalent to the bifurcation singularities of the zeros of a real polynomial of degree $l$
\cite{Arnold1986}. The liquid-glass transition points are $A_2$ singularities
located on smooth surfaces in
parameter space. If ${\mathbf V}$ crosses this surface at some ${\mathbf V}^c$, the 
long-time limit $\phi_q(t\rightarrow\infty)$ jumps from zero to the critical 
Debye-Waller factor $f_q^c>0$. There may exist surfaces of $A_2$ singularities within the 
glass. These describe iso-structural transitions from one amorphous state characterized by
$f_q>0$ to another state specified by a larger Debye-Waller factor $f_q^c>f_q$. For every 
$A_2$ singularity, a number $\lambda,\,0.5\leq\lambda < 1$, can be calculated . It is called 
the exponent parameter since it determines the various anomalous exponents entering the 
description of the slow dynamics for ${\mathbf V}$ near ${\mathbf V}^c$. The higher-order 
singularities, $A_l,\,l\geq 3$,  are the endpoints of the transition surfaces characterized 
by $\lambda=1$. In the following, some properties of $A_4$ singularities ${\mathbf V}^*$
shall be discussed. Interparticle interactions shall be considered consisting of a 
hard-sphere-repulsion core of diameter $d$ and a short-ranged attraction potential for $r>d$. The latter
shall be parameterized by a typical attraction strength $u_0$ and a typical attraction range 
$\Delta$. Therefore, a three-dimensional control-parameter space is considered:
${\mathbf V}=(\varphi,\Gamma,\delta)$, where $\varphi=(\pi/6)\rho d^3$
denotes the packing fraction of the spheres, $\Gamma=u_0/(k_{\rm B}T)$ is a dimensionless 
attraction strength or an inverse dimensionless temperature, and $\delta=\Delta/d$ is an 
attraction-range parameter.

For a system of particles interacting with a hard-sphere repulsion complemented by a short-ranged
attraction, MCT leads to two intriguing results \cite{Fabbian1999,Bergenholtz1999}. First,
the increase of the attraction strength $\Gamma$ may cause a melting of the glass. This implies
a reentry phenomenon for the transition diagram. For certain values of the packing fraction $\varphi$,
the liquid freezes into a glass not only by cooling but also by heating. This prediction has been
verified recently for colloidal suspensions \cite{Pham2002,Eckert2002}. In this work, the parameter
$\Gamma$ was varied by adding polymer to the solvent thereby increasing the strength of the 
depletion attraction. The reentry phenomenon was also established by molecular-dynamics-simulation
studies \cite{Pham2002,Puertas2002,Foffi2002prec,Zaccarelli2002preb}.
Second, the existence of an $A_3$ singularity was predicted. The glass-glass transitions connected
with this higher-order singularity deal with discontinuous changes of the localization mechanism
from one caused by the repulsion-induced cage effect to one caused by the attraction-induced bonding
of the particles. The cited work was based on Baxter's model for sticky hard spheres (SHS)
\cite{Baxter1968b} complemented with a large cutoff wave vector $q_{\rm co}$ restricting the 
wave-vector sums in the mode-coupling functional from \Eref{eq:Fdef}. The effective range of the
regularized potential introduced by the cutoff can be parameterized
by $\Delta=\pi/q_{\rm co}$. The singular Baxter interaction leads to 
a large-$q$-tail for the direct correlation function $c_q={\cal O}(1/q)$. It is this tail for
$2\pi/d<q<q_{\rm co}$ that causes the $A_3$ bifurcation. 
This $A_3$ singularity disappears if $q_{\rm co}$ is decreased towards $2\pi/d$.

The generic scenario for the disappearance of an $A_3$ singularity in a three parameter system is 
the existence of an $A_4$ glass-transition singularity at some point 
${\mathbf V}^*=(\varphi^*, \Gamma^*, \delta^*)$. Thus, the cited results 
\cite{Goetze1991b,Arnold1986,Fabbian1999,Bergenholtz1999} lead to the following conclusion. 
A system of particles interacting with some steep strong repulsion core complemented by a
sufficiently strong short-ranged attraction potential exhibits an $A_4$ singularity at some
control-parameter point ${\mathbf V}^*$. The general transition diagram is organized around this
${\mathbf V}^*$. Diagrams for different models for the interparticle interaction can be 
mapped onto each other for small ${\mathbf V}-{\mathbf V}^*$ by a smooth invertible parameter
transformation. In this sense, the bifurcation scenario is universal. The scenario at an
$A_4$ singularity has been demonstrated in some detail for the square-well system
\cite{Dawson2001}. In the following , these results shall be extended and
compared with the ones calculated for
other potentials that might be of interest for the description of colloidal systems.

\section{Results}\label{sec:result}

Let us specify the attraction potentials to be considered and the approximation theories to be used 
to evaluate $S_q$. The model of SHS complemented by
the mentioned cutoff $q_{\rm co}$ shall be used with the convention $\Gamma=15/\tau$. For the 
definition of the stickiness parameter $\tau$ and the evaluation of the structure factor we follow 
Baxter's work \cite{Baxter1968b}. The hard-core Yukawa system (HCY) is used with the following
convention for the control parameters 
\begin{equation}\label{eq:HCY}
U(r)/(k_{\rm B}T) = -\Gamma \exp[-(r-d)/(\delta d)]/(r/d)\,,\quad d<r\,,
\end{equation}
i.e., $\delta$ is chosen as the inverse of the conventional screening parameter $b$.
The structure factor is evaluated analytically in the mean-spherical approximation 
\cite{Cummings1979}.
Furthermore, three potentials of polynomial shape are considered:
\begin{equation}\label{eq:POLY}
U(r)/(k_{\rm B}T) = -\Gamma [(d+\delta d-r)/(\delta d)]^{(n-1)}\,,\quad d<r<(1+\delta)d\,.
\end{equation}
For $n=1$, the formula describes the square-well system (SWS). The triangular-potential model (TRI)
is obtained for $n=2$. The Asakura-Oosawa system (AOS) is modelled by $n=3$. The latter 
potential is obtained as small-$\delta$ limit for the depletion attraction acting between spheres
in a dilute solvent of small polymers \cite{Asakura1958}. It was shown that the equations for the 
mean-spherical approximation for $S_q$ of the SWS can be solved analytically by an 
expansion in $\delta$ \cite{Dawson2001}. This procedure can be extended to treat the potentials 
of \Eref{eq:POLY} for every $n$. The leading order expansion formulae are noted in \ref{sec:Sq}, 
and they are used in the following. Let us recall, that for the range of length-parameters
of interest, say $\delta<0.20$ for the SWS, the next-to-leading-order result is very close to the 
full numerical solution for $n=1$ \cite{Dawson2001}.

To determine $f_q$ for a given ${\mathbf V}$, the standard iteration procedure is applied:
$f_q^{(n)}/(1-f_q^{(n)}) = {\cal F}_q[f_k^{(n-1)}],\,n=1,2,\dots,\,f_q^{(0)}=1,\,
\lim_{n\rightarrow\infty}f_q^{(n)}=f_q$ \cite{Goetze1991b}. This is done after 
Equations~(\ref{eq:F},\ref{eq:vertex}) are rewritten so that the wave-vector moduli are discretized 
to $M$ points on a grid of equal spacing $h$. The values $h=0.4/d$ and $M=300$ have been used to
identify the $A_2$ singularities on the bifurcation surface. It is cumbersome to identify the
higher-order singularities since $(l-1)$ control parameters have to be scanned and the convergence
of the iteration is slower for points at an $A_{l+1}$ singularity than at an $A_{l}$ one. One can
use the deviation of $\lambda$ from unity to characterize the error for the identification of an 
$A_l$ with $l\geq 3$. In our calculations we achieved $1-\lambda<10^{-3}$. At the end, the calculation
at ${\mathbf V}^*$ was repeated with  $h=0.08/d$ and $M=1500$ in order to check the independence 
of the results from the discretization approximation.

\begin{figure}[htb]
\centerline{\includegraphics[width=0.8\textwidth]{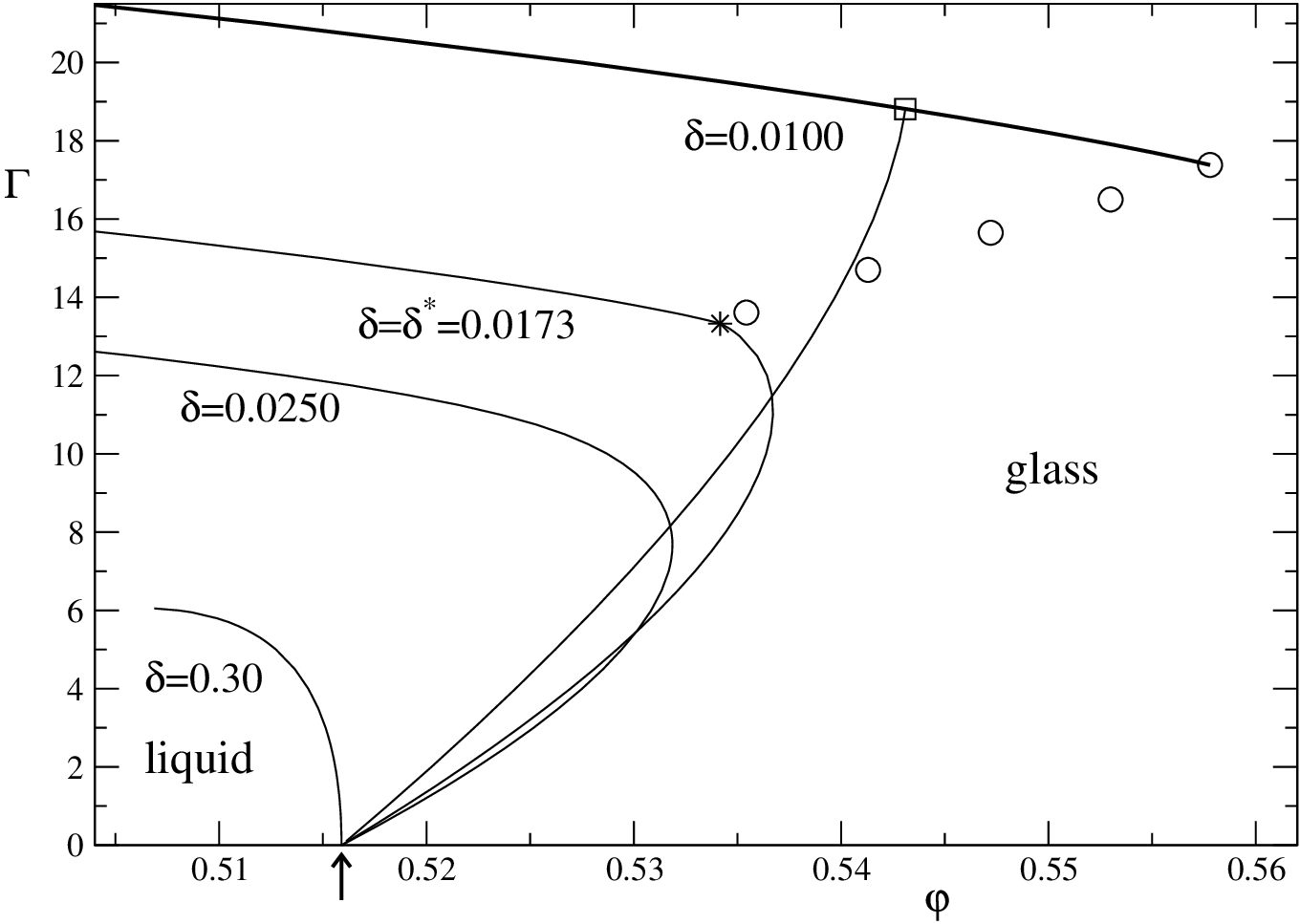}}
\caption{\label{fig:hcy_PD}
Transition diagram for the hard-core-Yukawa system (HCY). The attraction strength $\Gamma$ versus 
packing-fraction $\varphi$ curves show cuts through the surfaces of glass-transition singularities
for fixed attraction-range parameter $\delta$ as indicated; cf.\Eref{eq:HCY}.
The line for $\delta=\delta^*=0.0173$ hits the $A_4$ singularity ({\large$\ast$}). The line for 
$\delta=0.0100$ exhibits a crossing point ($\square$), and an $A_3$ singularity ($\bigcirc$). Four further
$A_3$ singularities are marked by circles; they refer from left to right to 
$\delta=0.0167,\,0.0143,\,0.0125,\,0.0111$. The arrow points to the critical packing fraction 
$\varphi^c_{\rm HSS}=0.516$ for the hard-sphere system. 
}
\end{figure}

As a representative example, \fref{fig:hcy_PD} exhibits the glass-transition-singularity diagram
of the HCY. The inner points of the lines are $A_2$ singularities obtained as constant-$\delta$ cuts
through the bifurcation manifold in the three-dimensional space of control parameters ${\mathbf V}$.
There are three topologically different cuts. The smooth
line shown for $\delta=0.0250$ represents a typical cut for $\delta>\delta^*$. All points describe
liquid-glass transitions characterized by an exponent parameter $\lambda<1$. For $\delta=\delta^*=
0.0173\dots$, the transition line runs through the $A_4$ singularity marked by a star. For all
${\mathbf V}\neq{\mathbf V}^*$, the line is smooth and  $\lambda<1$. For 
${\mathbf V}={\mathbf V}^*$, $\lambda=1$ and the line has a continuous tangent but an infinite 
curvature. The lines for $\delta<\delta^*$ consist of two pieces as demonstrated for 
$\delta=0.0100$. The part shown as a heavy line with $\rmd\Gamma^c/\rmd\varphi^c<0$ terminates within 
the glass in an $A_3$ singularity marked by a circle. The second part shown as a light line terminates
in a crossing point indicated by a square. The line of $A_2$ singularities between the crossing
point and the  $A_3$ singularity deals with glass-glass transitions and the other part with
liquid-glass transitions.
$A_2$-glass-transition curves for the HCY for $\delta\neq\delta^*$ cuts have been considered previously
for the HCY \cite{Bergenholtz1999,Foffi2002,Dawson2002}.

Let us add a remark on the diagram. Obviously, for $\Gamma=0$, all cuts start at the critical 
packing fraction of the hard-sphere system (HSS), $\varphi^c_{\rm HSS}=0.516$. There is a 
characteristic range parameter $\delta_{\rm reentry}$ so that the slope $\rmd\Gamma^c/\rmd\varphi^c$
for $\Gamma=0$ is negative for $\delta>\delta_{\rm reentry}$ and positive for 
$\delta<\delta_{\rm reentry}$. The latter case is exemplified in \fref{fig:hcy_PD} by the three 
cuts discussed in the preceding paragraph, the former case is shown by the cut for $\delta=0.300$.
For $\delta<\delta_{\rm reentry}$ and sufficiently small $\Gamma$, the transition curves deal with
melting of the glass upon increasing the attraction parameter $\Gamma$. Hence, for all liquid states
with $\varphi>\varphi^c_{\rm HSS}$, one gets the reentry phenomenon mentioned in the introduction.
The bonding forces create holes in the cages and this can destroy the particle localization for long 
times. In
agreement with Lindemann's melting criterion, the critical localization length of the HSS is about
$r_s^c=0.0746d$. If the attraction-potential range $\Delta$ is much larger than $r_s^c$, i.e., if $\delta$
exceeds a critical value $\delta_{\rm reentry}$, the 
bonding effects cannot change the cage structure. Therefore, the reentry mechanism disappears for
$\delta > \delta_{\rm reentry}$. In this case, the transition curve is similar to the one obtained for
a typical van-der-Waals system described, e.g., by a Lennard-Jones potential. One obtains 
$\delta_{\rm reentry}=0.30$ and $0.117$ for HCY and SWS, respectively. It should be noted that the 
transition curve shown for $\delta=0.30$ terminates at the spinodal line defined by the divergence of $S_q$
for $q=0$. Beyond this point, MCT equations are meaningless and calculations of density correlators cannot 
be performed there within that theory.

\begin{figure}[htb]
\centerline{\includegraphics[width=0.8\textwidth]{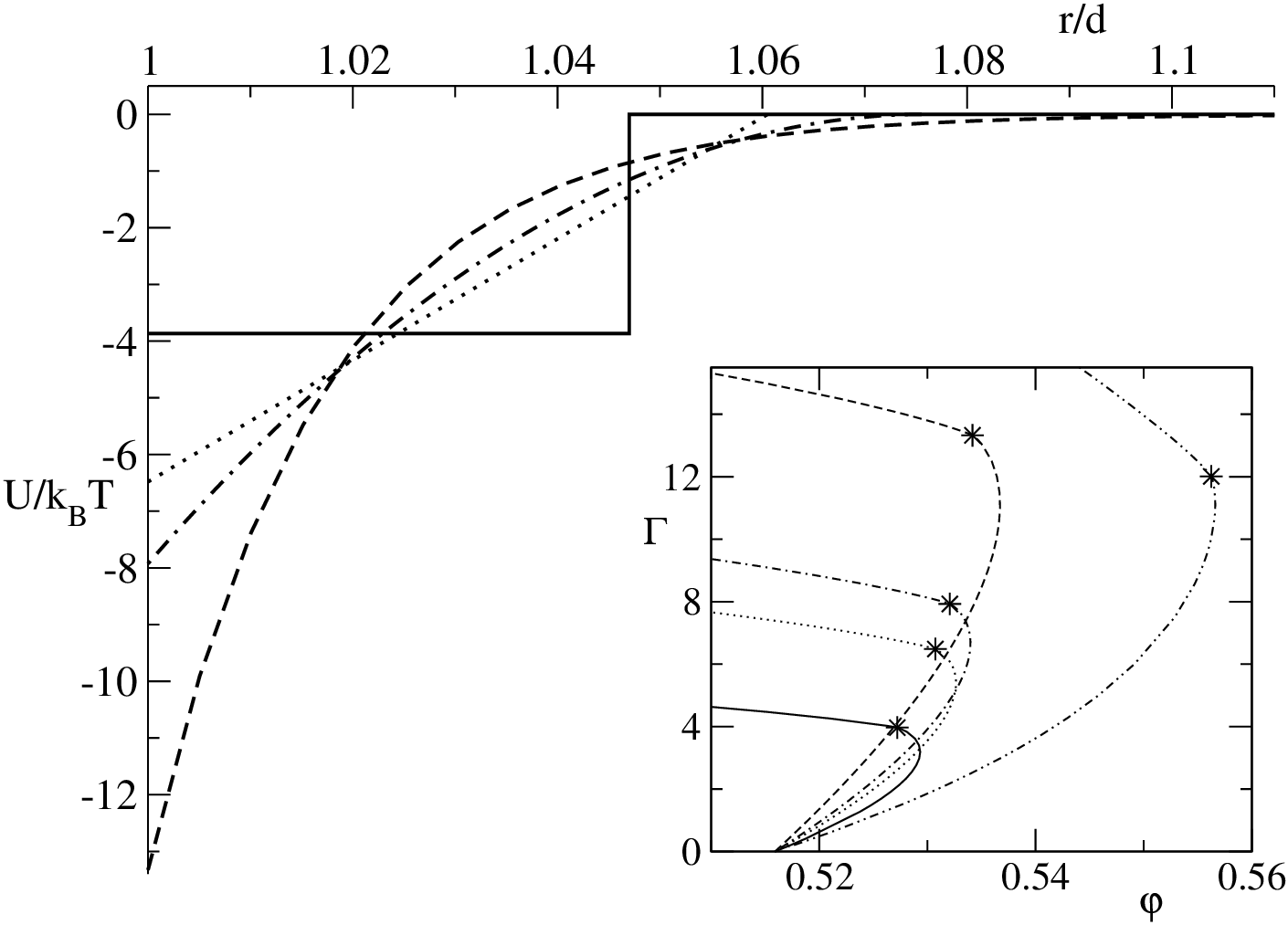}}
\caption{\label{fig:pot}
Attractive potentials $U$ relative to the thermal energy $k_{\rm B}T$ as
function of the interparticle distance $r$ relative to the hard-sphere diameter $d$
for control parameters $V^*$ at the respective $A_4$ singularity. The curves refer to the 
SWS~(---), TRI~($\cdots$), AOS~($-\cdot-$), HCY~($--$) and SHS~($-\cdot\cdot-\cdot\cdot-$).
The inset shows the cuts of the bifurcation surfaces through ${\mathbf V}^*$ for fixed
attraction-range parameter $\delta=\delta^*$.
}
\end{figure}

\Fref{fig:pot} exhibits the four regular attraction potentials defined in 
Equations~(\ref{eq:HCY},\ref{eq:POLY}) for control parameters at the respective $A_4$ singularities.
The inset shows cuts through the liquid-glass-transition surfaces for fixed range parameter $\delta^*$.
The regular potentials are rather close to each other for $1.01 \leq r/d \leq 1.05$. The volume of the 
shell $1.00<r/d<1.02$ is smaller than the one for the shell $1.02<r/d<1.04$. Within the latter, the
attraction strength decreases in the sequence SWS, TRI, AOS and HCY. Therefore, the critical packing 
fraction $\varphi^*$ for the onset of an iso-structural glass transition increases in this sequence. One 
finds $\varphi^* = 0.5272,\,0.5307,\,0.5321,$ and $0.5342$, respectively. For the same reason, the 
maximum packing fraction of the liquid increases in this sequence. One gets 
$\varphi^{\rm max} = 0.5293,\,0.5326,\,0.5340,$ and $0.5367$, respectively. For Baxter's model, all
mode-coupling effects have been cut off for $q>\pi d/\delta^*$. Therefore, the bonding effects are
reduced compared to those for the regular potentials and the values $\varphi^* = 0.5562$ and 
$\varphi^{\rm max} = 0.5566$ for the SHS are larger than the corresponding values for the regular potentials.
Evaluating the structure factor of the SWS up to next-to-leading order as done in Reference~\cite{Dawson2001}, 
one gets $\varphi^* = 0.5277$ and $\varphi^{\rm max} = 0.5299$. Indeed, the difference of these numbers from 
those calculated with the formulae from \ref{sec:Sq} is small.

\begin{figure}[htb]
\centerline{\includegraphics[width=0.8\textwidth]{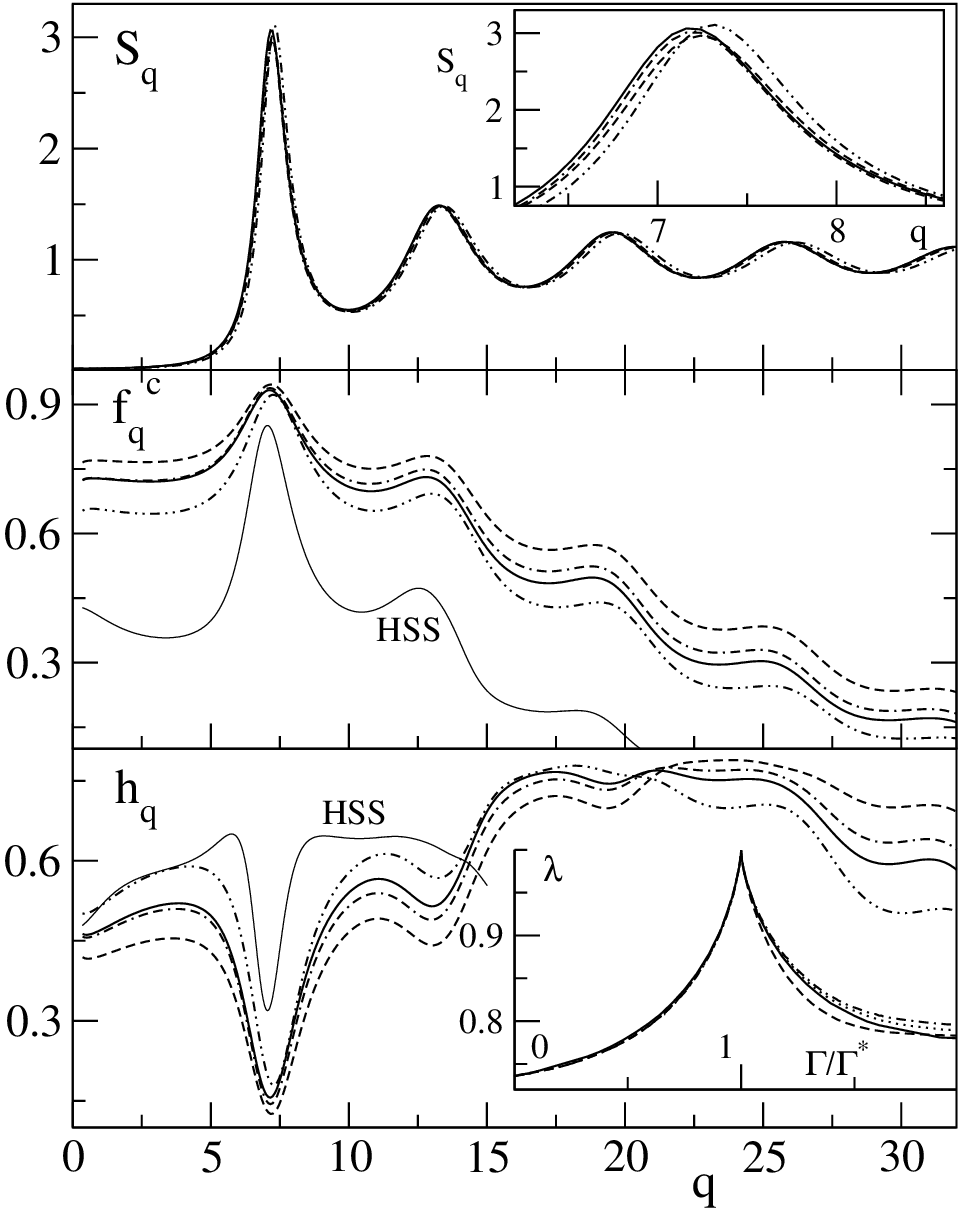}}
\caption{\label{fig:A4}
Structure factor $S_q$, critical Debye-Waller factor $f_q^c$, and critical amplitude $h_q$
for control parameters specifying the $A_4$ singularity $V^*$, SWS~(---), HCY~($--$) and 
SHS~($-\cdot\cdot-\cdot\cdot-$). The fourth curve ($--\cdot--\cdot$) shows the results for the SWS using the 
structure-factor up to next-to-leading order in $\delta$ \cite{Dawson2001}. 
The lines denoted by HSS exhibit $f_q^c$ and $h_q$
for the critical point of the hard-sphere system. The inset from the lower panel shows the variation of 
the exponent parameter $\lambda$ along the liquid-glass transition lines through ${\mathbf V}^*$ from 
\Fref{fig:pot}.
}
\end{figure}

Structure factors $S_q$ at the $A_4$ singularities and the corresponding critical Debye-Waller factors
$f_q^c$ are compared in \Fref{fig:A4}. Also compared are the so-called critical amplitudes $h_q$.
These are evaluated by a straightforward but involved procedure from the mode-coupling functional
at the bifurcation point \cite{Goetze1991b}. They quantify the susceptibility of the arrested structure.
The increase of $(f_q-f_q^c)$ upon increasing ${\mathbf V}-{\mathbf V}^*$ is proportional to $h_q$ and 
so is the prefactor in the logarithmic decay laws that are the characteristic feature of the dynamics 
near a higher-order singularity \cite{Goetze2002}. $S_q$ for various models mainly differ by a small shift 
parallel to the $q$-axis only. This shift reflects the decrease of the interparticle distance caused by the 
increase of $\varphi^*$. The $f_q^c$ oscillate in phase with $S_q$ and the $h_q$ oscillate in opposite phase 
as known and explained for the simple HSS \cite{Goetze1991b}. However, the $f_q^c$ are considerably larger and 
the $h_q$ are smaller than 
the corresponding quantities at the hard-sphere transition. This is a manifestation of the fact that the
attractive part of the potential enforces localization, provided a glass state is established. Within
the wave-vector region around the first diffraction peak, say $1\leq qd\leq 10$, the Debye-Waller factor
of the SWS differs from the one of the HCY up to about 7\% while this difference is minimal at the peak.
The corresponding difference for the critical amplitude is about 9\% which is maximal at the peak.
The inset for the lowest panel of \Fref{fig:A4} displays the variation of the exponent parameter 
$\lambda$ for the liquid-glass transitions on the cuts $\delta=\delta^*$. Rescaled as a function
of $\Gamma/\Gamma^*$, the $\lambda$ cannot be distinguished on the branch $\Gamma/\Gamma^*<1$ dealing
with transitions to the repulsion dominated glass. On the branch $\Gamma/\Gamma^*>1$, which deals with
transitions to the attraction dominated glass, the $\lambda$ for the various models are still very close
to each other.

\Fref{fig:A4} also exhibits results for the SWS evaluated with the structure factor in next-to-leading order
of Reference~\cite{Dawson2001} in comparison with those based on the leading order theory explained in 
\ref{sec:Sq}. Obviously the difference is too small to be of interest.

The macroscopic mechanical stiffness of liquids and glasses is quantified by the elastic moduli. The
longitudinal modulus $M_{\rm L}$ specifies the stiffness for compressions and the transversal one $M_{\rm T}$,
also called shear modulus $G'$ , the stiffness for shear deformations. They are defined as constants of 
porportionality in the linearized stress-strain relation. In systems with Newtonian microscopic dynamics,
they determine the speed of longitudinal and transversal sound, respectively, via 
$v_{\rm L,T} = \sqrt{M_{\rm L,T}/(\rho m)}$ with $\rho m$ denoting the mass density. For an ergodic system,
the shear modulus vanishes, $M_{\rm T}^0=0$. The longitudinal modulus reads $M_{\rm L}^0=\rho(k_{\rm B}T)S_0^{-1}$
with $S_0=\lim_{q\rightarrow 0}S_q$. In the glass state, the moduli are larger: 
$M_{\rm L,T}=M_{\rm L,T}^0 + \delta M_{\rm L,T}$. The additional contributions are the positive long-time 
limits of fluctuating-force correlators. For the latter, MCT yields \cite{Goetze1991b}:
\begin{equation}\label{eq:deltaG}
\delta M_{\rm L,T} = \rho (k_{\rm B}T) \lim_{q\rightarrow 0}\sum_{\vec{k}+\vec{p}=\vec{q}}
S_k S_p f_k f_p \{[\vec{k}c_k+\vec{p}c_p]\vec{e}_{\vec{q}}^{\rm L,T}\}^2 (\rho/2 q^2) \,.
\end{equation}
Here $\vec{e}_{\vec{q}}^{\rm L,T}$ are unit vectors parallel and perpendicular to $\vec{q}$, respectively. 
The limit leads to
\numparts\label{eq:M_result}
\begin{eqnarray}\label{eq:M_result:a}
\delta M_{\rm L,T} &=& \rho (k_{\rm B}T)\int_0^\infty\rmd k \{\rho [S_k f_k k/(2\pi)]^2 w_{\rm L,T}(k)\}\,,\\
\label{eq:M_result:b}
w_{\rm L}(k) &=& c_k^2 + \frac{2}{3}(kc_k')c_k+\frac{1}{5}(kc_k')^2\,,\\\label{eq:M_result:c}
w_{\rm T}(k) &=& \frac{1}{15}(kc_k')^2\,.
\end{eqnarray}
\endnumparts

\begin{figure}[htb]
\centerline{\includegraphics[width=0.8\textwidth]{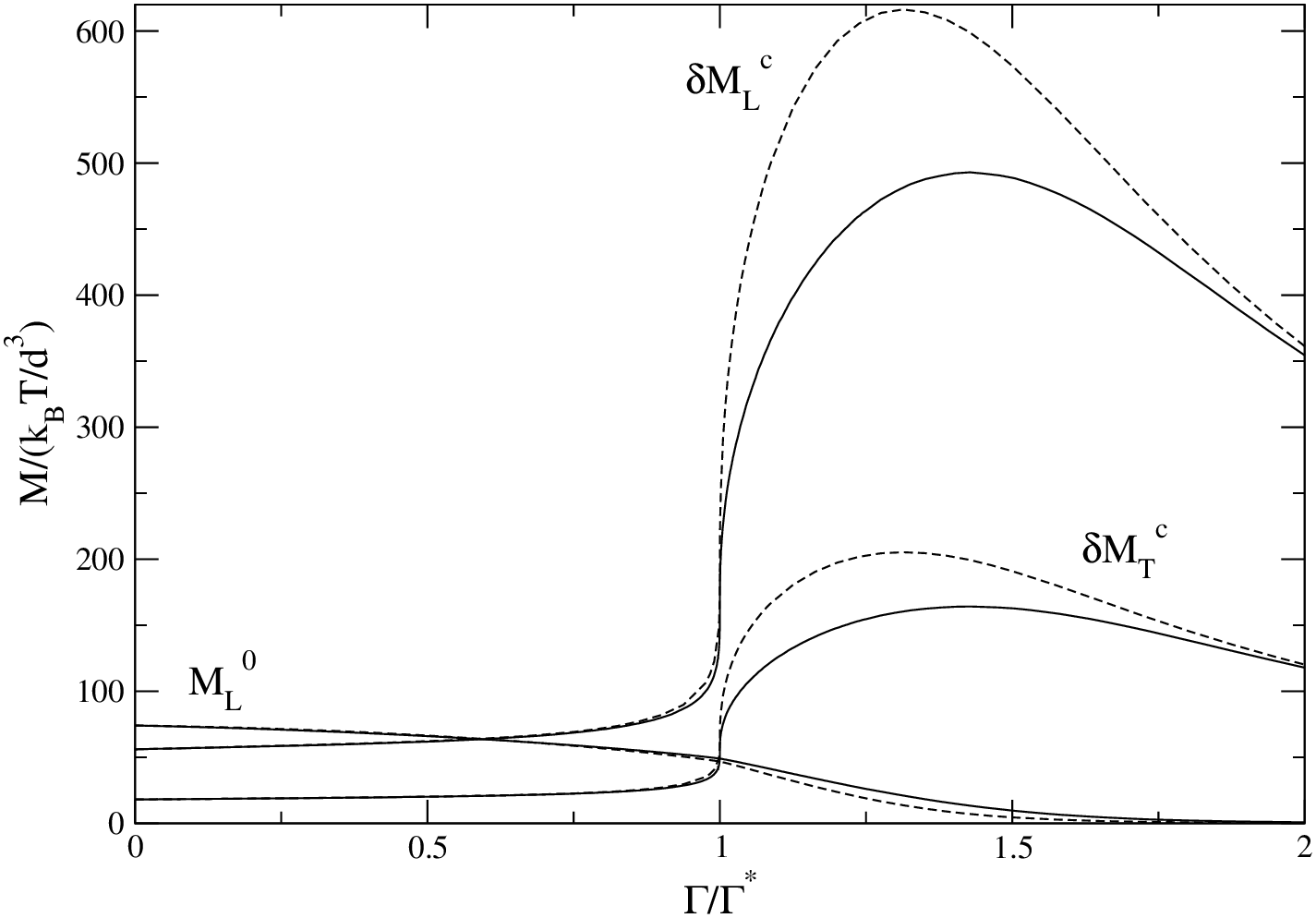}}
\caption{\label{fig:moduli}Longitudinal elastic moduli $M_{\rm L}$ and transversal elastic moduli $M_{\rm T}$
for the $\delta=\delta^*$ transition line
of the SWS (HCY) shown as full (dashed) lines. For the SWS, the 
structure factor from \cite{Dawson2001} was used. In the limiting case of the HSS, 
$M_{\rm L}^0=74.1$, $\delta M_{\rm T}=18.1$, $\delta M_{\rm L}=56.1$. At the $A_4$ singularity for the SWS (HCY), 
$\delta M_{\rm T}^*=  54.3(64.4)$, $\delta M_{\rm L}^*= 163.4 (193.6)$.
At the maximum $\delta M_{\rm T}= 164.2 (205.3)$, $\delta M_{\rm L}= 493.0 (616.3)$.
}
\end{figure}

In \Fref{fig:moduli} the moduli for the SWS are compared with the ones for the HCY. The states are chosen
on the cuts $\delta=\delta^*$ through the respective transition surface. The compression modulus $M^0$ of the liquid 
varies smoothly throughout reflecting the well known increase of the compressibility with increasing 
attraction forces. The large variations of 
$\delta M_{\rm L,T}$ reflect the strong effect of bonding potentials on restoring forces 
\cite{Bergenholtz1999,Dawson2001}. The same bonds are resisting shear as well as compression deformations. 
Therefore, there is no great difference in the behaviour of the two moduli. The contributions to $\delta M_{\rm L}$
due to the first two terms in \Eref{eq:M_result:b} are smaller than the one due to the last term. Further, 
incidentally, these two contributions nearly cancel. Therefore, $\delta M_{\rm L}^c$ differs from 
$3\delta M_{\rm T}^c$ by less than 3\% for $\Gamma<\Gamma^*$ and less than 0.5\% for $\Gamma\geq\Gamma^*$.
The universal properties of the 
$A_4$ bifurcation imply the following: The moduli vary smoothly for all $\Gamma\neq\Gamma^*$ and at the $A_4$
singularity there is a cubic-root singularity: $M(\Gamma)-M(\Gamma^*)\propto (\Gamma-\Gamma^*)^{1/3}$. The singular
increase of $M$ with $\Gamma$ increasing through $\Gamma^*$ is a precursor of the discontinuous increase of $M$
upon crossing the glass-glass-transition line for $\delta<\delta^*$. If $\Gamma$ increases further, the moduli 
have to decrease since the density decreases towards zero. Therefore, the moduli exhibit a maximum at some value
of $\Gamma$ exceeding $\Gamma^*$. By continuity, such maximum also occurs for cuts with $\delta$ close but not
equal to $\delta^*$, as was noticed before for the shear modulus \cite{Bergenholtz1999,Zaccarelli2001}.

\begin{figure}[htb]
\centerline{\includegraphics[width=0.8\textwidth]{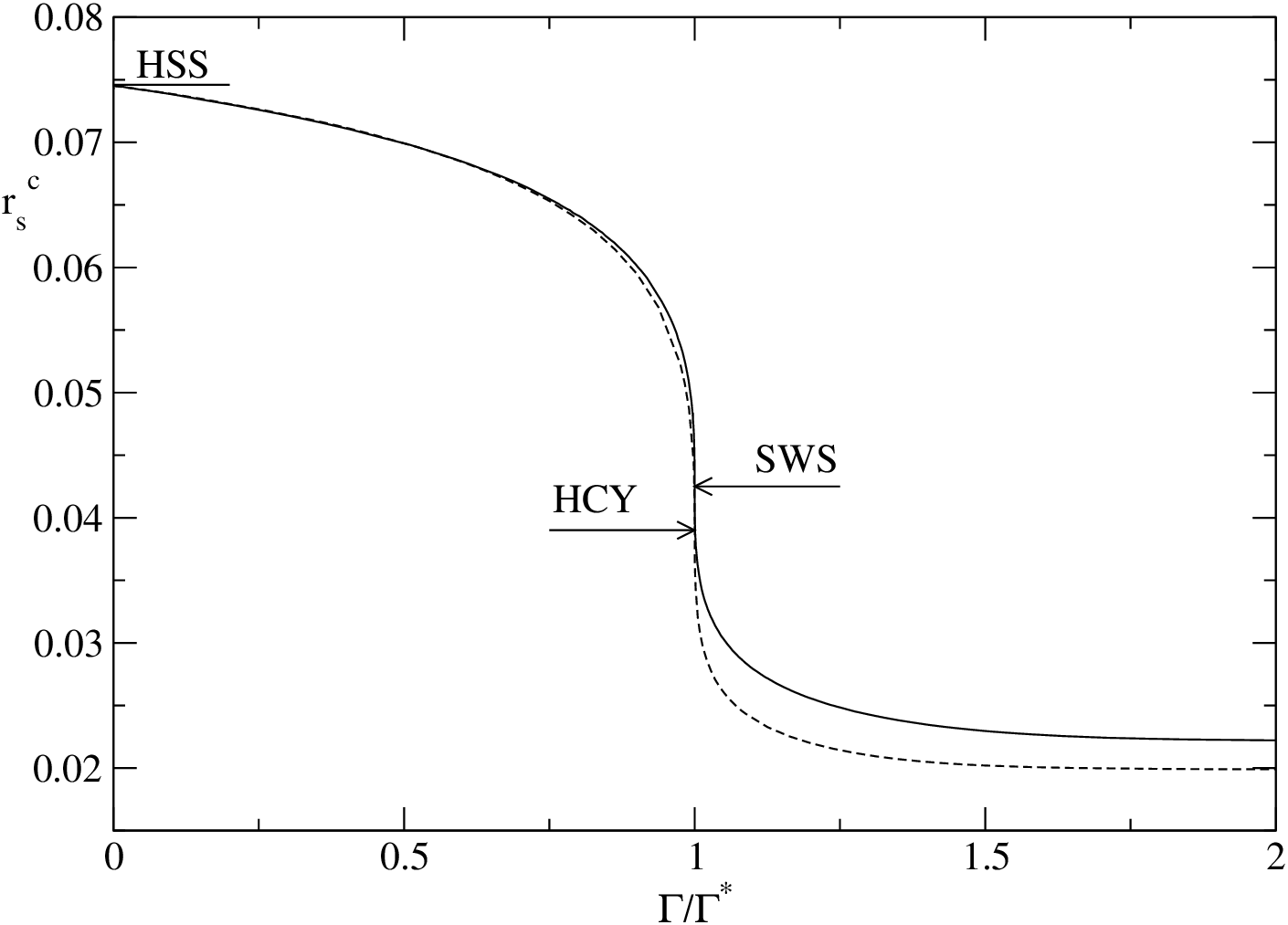}}
\caption{\label{fig:rsc}Localization length $r_s^c$ of a tagged particle for states on the
 $\delta=\delta^*$ transition line for the SWS (HCY) shown as full (dashed) line.  At the $A_4$ singularity 
one gets $r_s^c/d = 0.0425 (0.0390)$ as indicated by the arrows. The line labeled HSS marks the localization 
length  $r_s^c/d = 0.0746$ for the HSS.
}
\end{figure}

The clearest quantity for demonstrating the change of the glassification mechanism upon increasing the attraction,
is the variation of the particle's localization length $r_s$ as a function of $\Gamma$ for states on the cut
$\delta=\delta^*$ (\Fref{fig:rsc}). This length is given by the mean-squared radius of a particle's 
long-time probability density. The Fourier-transform of the latter is the Lamb-M\"o\ss bauer factor $f_q^s$ that
is evaluated from an equation similar to \Eref{eq:F} \cite{Goetze1991b}. One ends up with an equation for the 
inverse of $r_s^2$ analogous to the equations for the moduli:
\begin{equation}\label{eq:rsc}
1/r_s^2 = \frac{1}{6}\int_0^\infty\rmd k \{\rho S_k f_k f_k^s [c_k k^2/\pi]^2\}\,.
\end{equation}
With increasing $\Gamma$, $r_s^c$ decreases monotonously from the value for the HSS, $r_s^c/d = 0.0746$. The 
variation is smooth except for the cubic-root sigularity at $\Gamma = \Gamma^*$. For strong bonding, $r_s^c$
decreases to a $\Gamma$ insensitive value of the order of the attraction potential range, as noted before
for $\delta>\delta^*$ in the HCY \cite{Bergenholtz1999}.

\section{Discussion}\label{sec:discuss}

The first peculiarity of structural relaxation in systems of particles interacting with hard-sphere-like
repulsions complemented by short-ranged attractions predicted by mode-coupling theory (MCT) 
\cite{Fabbian1999,Bergenholtz1999}, namley the reentry phenomenon for the liquid-glass-transition diagram, 
has been established recently \cite{Pham2002,Eckert2002,Puertas2002,Foffi2002prec,Zaccarelli2002preb}.
This justifies to focus now on the more difficult
task of testing the second prediction concerning the existence of $A_3$ and $A_4$ singularities. The
signature of such higher-order singularities is the extreme stretching of relaxation curves as has been
explained on the basis of asymptotic expansions in terms of powers of logarithms in time 
\cite{Goetze2002}. The recent simulation results \cite{Puertas2002,Zaccarelli2002preb} provide hints that
there are such logarithmic decay laws. However, a more detailed analysis of the data would be required 
if one intends to arrive at compelling conclusions.
The transition diagram exhibiting $A_3$ singularities is organized around an $A_4$ singularity. So far, such
singularity was identified within a microscopic
theory only for the square-well system \cite{Dawson2001}. For this
system, all parameters and amplitudes necessary for a discussion of the logarithmic decay laws
have been evaluated \cite{Sperl2002}. In the present paper, $A_4$ singularities have been identified and
characterized for a series of other attraction potentials in order to provide some information on the
sensitivity of the MCT results on the microscopic details assumed for the interaction.

The critical packing fractions $\varphi^*$ found for the $A_4$ singularities of the
four regular potentials considered 
(\Fref{fig:pot}) or the related largest values for the liquid packing $\varphi^{\rm max}$ exceed the critical
packing fraction $\varphi^c_{\rm HSS}$ of the hard-sphere system by about 3\%. But the difference between 
these values for different models
is about 1.4\% only. It is not clear, whether or not experiments can discriminate between 
the $\varphi^*$ or $\varphi^{\rm max}$ predicted for, say, SWS and HCY. But simulation studies for the
diffusivity of the kind published recently \cite{Foffi2002prec,Zaccarelli2002preb} should be able to test our
prediction $\varphi^{\rm max}_{\rm HCY}>\varphi^{\rm max}_{\rm SWS}$.

The singular potential for sticky hard spheres was introduced by Baxter \cite{Baxter1968b} as a certain
attraction-range-to-zero limit in order to simplify equations from a mathematical point of view. But, 
within MCT, these simplifications cannot be made use of since the limit to zero for the range parameter
$\delta$ must not be permuted with the limit to infinity for the wave vectors in the mode-coupling 
integrals. The problem can be handled by introducing some wave-vector cutoff 
\cite{Fabbian1999,Bergenholtz1999}. This is equivalent to introducing a regular potential leading to a
considerably larger $\varphi^*-\varphi^c_{\rm HSS}$ than found for the other potentials examined above.

It is well known that the structure factor theories used in this paper lead to unsatisfactory results for 
the thermodynamic functions. Great progress has been made to improve on this point as is discussed, e.g., 
in Reference~\cite{Foffi2002}. The thermodynamic functions are derived from the zero-wave-vector limits
of correlators. However, different from the limit of large wave vectors, which is important to capture the 
essential physics
of the glass transition, small-$q$ effects play a minor role for the transition curves.
As already demonstrated for the SWS in \cite{Dawson2001}, artificially setting the structure factor input to zero 
for $qd<4$ does not alter the transitions qualitatively. For $\Gamma\gtrapprox \Gamma^*$ where large $q$
values dominate, transition lines are in accordance even quantitatively.

The differences among the structure factors of the various models for wave vectors accessible in 
scattering experiments for colloids, say $2\leq qd \leq 15$, are very small if the states at the $A_4$ singularity
are compared (\Fref{fig:A4}). However, the differences between the Debye-Waller factors and the critical
amplitudes are of the order of 5\%. The $f_q^c$ and $h_q$ determine the amplitudes of the leading-order
formulae for the intermediate scattering functions for states near an $A_4$ singularity in a similar 
manner as they determine the corresponding results near an $A_2$ liquid-glass transition 
\cite{Goetze2002}. If they could be deduced from a data analysis with similar accuracy as one can do 
from data near an $A_2$ singularity, one could discriminate, e.g., between the square-well model and the
hard-core-Yukawa system.

The tagged particle localization is described by the Lamb-M\"o\ss bauer factor, $f_q^s$, that can be measured 
by incoherent scattering. It is a bell-shaped function of $q$ and its width can be quantified by the 
localization length $r_s$: $f_q^s = 1-(q r_s)^2+{\cal O}(q^4)$. As expected from the shape of the attractive
potentials at the $A_4$ singularity (\Fref{fig:pot}), $r_s^*$ is smaller for the HCY than for the SWS by 10\%.
The MCT brings out that $r_s^c$ becomes insensitive to $\Gamma$ for strong attraction. For the SWS (HCY) one 
gets $r_s^c/d=0.022$ ($0.020$) for $\Gamma/\Gamma^*>1.8$ within 1\% accuracy.

The largest difference between SWS and HCY is found at the maximum of the elastic moduli where the HCY is
about 20\% stiffer than the SWS. Taking the critical moduli of the HSS as reference, the maxima of SWS
and HCY should be at 9 or 11 times that value, respectively, as shown in \Fref{fig:moduli}.

The line of glass-glass transitions occurring for the cuts through the bifurcation surface for 
$\delta<\delta^*$ has some analogue in the line of iso-structural phase transitions from one face-centered 
cubic crystal to another one with a different lattice constant identified for systems similar to the ones 
discussed above \cite{Bolhuis1994a,Bolhuis1994,Tejero1994,Tejero1995}. 
Similar to what is shown in \Fref{fig:hcy_PD}, also the endpoint of the phase-transition line for the HCY 
exhibits a critical attraction constant $\Gamma^c$ that increases with decreasing range parameter $\delta$. 
However, the value $\Gamma^c$ for phase transitions of the SWS was found practically independent of $\delta$ 
\cite{Bolhuis1994}. Contrary to this finding for the crystal-crystal transition, the endpoint of the
glass-glass transition line of the SWS behaves quite similar to what is found for the HCY.

Summarizing, one can conclude that the MCT predictions for the dynamics near an $A_4$ singularity are rather 
robust. At the present state of discussion, one can use any of the so far studied models to predict data
semiquantitatively. Two reservations are required. First, the quality of the theories applied 
for the evaluation of the structure factors $S_q$ is not known for the high-density regime of interest. The
$S_q$ are, however, the essential input functions for all quantitative considerations. Second, the range of 
validity of the basic MCT equations of motion are not understood, not even qualitatively. Quantitative 
results of MCT have been tested with encouraging outcome for the hard-sphere system \cite{Megen1995}, 
binary Lennard-Jones systems \cite{Nauroth1997}, and silica \cite{Sciortino2001}.
It remains to be tested by experiment or computer simulation whether or not the theory can also describe 
the systems studied in this paper.

\ack
We thank J.~Bergenholtz, M.~Fuchs and Th.~Voigtmann for discussion and J.~Bergenholtz for assistance 
while implementing the HCY structure factor. Our work was supported by the DFG grant Go~154/13-1.

\appendix
\section{Structure Factors}\label{sec:Sq}
For the calculation of the structure factors we use a scheme developed earlier for the SWS \cite{Dawson2001}. 
The short-ranged attraction of the potential from \Eref{eq:POLY} added to the hard core is treated in 
mean-spherical approximation and the resulting equations are expanded in the small parameter $\delta$. 
The structure factor is expressed in terms of Baxter's factor function $Q(r)$ \cite{Hansen1986},
\begin{eqnarray}\label{eq:Qr_def}
S_q^{-1} &=& \hat{Q}(q) \hat{Q}(q)^*\,,\\
\hat{Q}(q) &=& 1 -2 \pi\rho \int_0^\infty \rmd r \exp[iqr] Q(r)\,.
\end{eqnarray}
In the leading-order approximation, one gets a shifted parabola within the hard core,
$0\leq r \leq 1$, 
\begin{equation}\label{eq:Qr_result_HS}
Q(r) = \frac{a}{2} (r^2-1) + b (r-1) + \frac{K}{n}\,,\\ 
\end{equation}
and a polynomial within the attraction shell, $1\leq r \leq (1+\delta)$:
\begin{equation}\label{eq:Qr_result_add}
Q(r) = \frac{K}{n}\left(\frac{1+\delta -r}{\delta }\right)^{n} \,,\quad 
1\leq r \leq 1+\delta\,.
\end{equation}
Here, the hard-core diameter $d$ is used as unit of length. For $r>(1+\delta)$, the factor 
function is zero. The various constants abbreviate $K=\Gamma\delta$,
\begin{eqnarray}\label{eq:ab_result}
a = \frac{1+2\varphi}{(1-\varphi)^2} -\frac{12\varphi}{(1-\varphi)}\frac{K}{n}\,, \quad
& b = \frac{-3\varphi}{2(1-\varphi)^2}+ \frac{6\varphi}{(1-\varphi)}\frac{K}{n}\,.
\end{eqnarray}
The limit $\delta\rightarrow 0$ can be carried out in 
Equations~(\ref{eq:Qr_result_HS}-\ref{eq:ab_result}) in the sense of 
Reference~\cite{Baxter1968b}. It yields the result for $S_q$ used in 
\cite{Fabbian1999,Bergenholtz1999}. However, this procedure appears unjustified.
Within the mode-coupling functional, \Eref{eq:Fdef}, one needs values of $S_q$ also for large
wave-vectors. But the limits $q\rightarrow\infty$ and $\delta\rightarrow 0$ cannot be permuted. 
Introduction of the cutoff is one way to redefine the model so that the MCT remains meaningful. A more
appealing procedure is to consider the $\delta$ expansion from Reference~\cite{Dawson2001} and 
define as Baxter limit the one where terms of order $\delta^1$ are neglected compared to those of
order $\delta^0$. But this is just the approximation defined by 
Equations~(\ref{eq:Qr_result_HS}-\ref{eq:ab_result}) that was used in this paper.
It is in no respect more complicated to handle than the approximation proposed originally \cite{Baxter1968b}.

\section*{References}

\end{document}